\documentstyle[floats,tighten,preprint,prl,aps,aps10,epsf]{revtex}

\begin{document}

% \twocolumn[\hsize\textwidth\columnwidth\hsize\csname
% @twocolumnfalse\endcsname

\preprint{\parbox[t]{15em}{\raggedleft
FERMILAB-Pub-00/148-T \\ hep-ph/0006345 \\[2.0em]}}
\draft

% add words to TeX's hyphenation exception list
\hyphenation{author another created financial paper re-commend-ed}

% declarations for front matter
\title{Computation of $\bar{\Lambda}$ and $\lambda_1$ 
with Lattice QCD}

\author{Andreas S. Kronfeld and James N. Simone}
\address{Fermi National Accelerator Laboratory, 
P.O. Box 500, Batavia, Illinois}

\date{July 31, 2000}

\maketitle % typeset front matter (abstract goes after for REVTeX)

% \narrowtext
% \widetext
\begin{abstract}
We pursue a new method, based on lattice QCD, for determining the
quantities $\bar{\Lambda}$, $\lambda_1$, and $\lambda_2$ of heavy-quark
effective theory.
We combine Monte Carlo data for the meson mass spectrum with
perturbative calculations of the short-distance behavior, to
extract~$\bar{\Lambda}$ and $\lambda_1$ from a formula from HQET.
Taking into account uncertainties from fitting the mass dependence
and from taking the continuum limit, we find
$\bar{\Lambda} = 0.68^{+0.02}_{-0.12}~\text{GeV}$
and $\lambda_1 = -(0.45 \pm 0.12)~\text{GeV}^2$
in the quenched approximation.
\end{abstract}

\pacs{PACS numbers: 12.38.Gc, 12.39.Hg, 13.20.-v}
% ]
% \narrowtext
\epsfverbosetrue
\twocolumn

In the past decade or so, heavy-quark effective theory (HQET) has 
become an indispensible tool for studying the physics of hadrons, such 
as $B$ and $D$ mesons, containing a single heavy quark.
The main physical idea is simple: as the heavy-quark mass increases, 
the wave function of a ``heavy-light'' hadron depends less and less on 
the heavy-quark mass~\cite{Shuryak:1980pg,Eichten:1990zv,Isgur:1989vq}.
This is precisely as in atomic physics, where properties of hydrogen 
and deuterium are almost the same.

A central result from HQET is the heavy-quark expansion of a hadron's 
mass.
Through order $1/m$, the mass~$M$ of a spin-$J$ meson ($J=0$, 1) 
is~\cite{Falk:1993wt}
\begin{equation}
	M = m + \bar{\Lambda}
		- \frac{\lambda_1}{2m} - d_J\frac{z_{\cal B}\lambda_2}{2m}
		+ O(1/m^2),
	\label{eq:M}
\end{equation}
where $d_0=3$ and $d_1=-1$ tracks the spin dependence.
Each term in Eq.~(\ref{eq:M}) has a simple physical interpretation:
$m$~is the heavy-quark mass, the definition of which is elaborated 
below;
$\bar{\Lambda}$~is the energy of the light quark and gluons; 
$-\lambda_1/2m$ is the kinetic energy of the heavy quark; 
and $d_Jz_{\cal B}\lambda_2/2m$ is the hyperfine energy of the
heavy quark's spin interacting with the chromomagnetic field inside
the meson.
The quantities~$\bar{\Lambda}$, $\lambda_1$, and~$\lambda_2$ in
Eq.~(\ref{eq:M}) describe the long-distance part of the bound-state
problem.
At long distances QCD is intrinsically nonperturbative, so it is not 
easy to calculate them from first principles.
This should be possible with lattice gauge theory, and the aim of this 
Letter is to demonstrate a new method for computing~$\bar{\Lambda}$, 
$\lambda_1$, and~$\lambda_2$.

Part of the utility of HQET is that the lambdas---$\bar{\Lambda}$,
$\lambda_1$, and~$\lambda_2$---appear also in
the heavy-quark expansions of inclusive decay
spectra~\cite{Chay:1990da,Bigi:1993fe,Blok:1994va,Falk:1994dh}.
Thus, they enter into the determination of the 
Cabibbo-Kobayashi-Maskawa (CKM) matrix elements 
$|V_{cb}|$~\cite{Luke:1994za,Ball:1994je}, and 
$|V_{ub}|$~\cite{Falk:1997gj,Bauer:2000xf}.
The spin splitting $M_{B^*}-M_B$ gives a simple way to 
estimate~$\lambda_2$, but meson masses alone are not 
enough to deduce~$\bar{\Lambda}$ and~$\lambda_1$.
Moments of inclusive decay
distributions~\cite{Gremm:1996yn,Gremm:1997gg,Falk:1998jq} 
do offer a way to relate experimental data to~$\bar{\Lambda}$ 
and~$\lambda_1$, but, nevertheless, an \emph{ab initio} QCD 
calculation should be of interest.

Before explaining our method for computing the lambdas, it is useful
to recall how they are defined.
HQET is an effective field theory, so it introduces an energy 
scale~$\mu$ to separate long- and short-distance physics.
All quantities (except~$d_J$) on the right-hand side of
Eq.~(\ref{eq:M}) depend on~$\mu$ and the renormalization scheme used
to define it.
(Meson masses remain independent of~$\mu$.)
Physics from distances shorter than~$\mu^{-1}$ is lumped into 
Wilson coefficients, such as $m$, $1/2m$ and $z_{\cal B}/2m$ in 
Eq.~(\ref{eq:M}).
Physics from distances longer than~$\mu^{-1}$ is described by operators 
in the Lagrangian of HQET.
The lambdas are matrix elements of these operators.
When computing them one should renormalize the operators so that the
lambdas are portable to the phenomenology of inclusive decays.
Because those analyses compute the Wilson coefficients in perturbative 
QCD, it is most common to renormalize HQET in a mass-independent scheme.
Then the quark mass~$m$ in Eq.~(\ref{eq:M}) is the pole mass of the 
underlying theory, i.e.,~QCD.
This choice of scheme obscures the $\mu$-dependent character of~$m$ 
and, thus, $\bar{\Lambda}$ and $\lambda_1$, but one should still think 
of the pole mass as a special choice of perturbative short-distance mass.
The scheme is easily portable, because the pole mass is infrared
finite and gauge independent at every order in perturbative
QCD~\cite{Kronfeld:1998di}, and the relation between the pole and
$\overline{\rm MS}$ masses in QCD is known through order
$\alpha_s^3$~\cite{Melnikov:2000qh}.

Another property of the lambdas is that they are independent of the 
heavy-quark mass (if, as we do, one distinguishes $\mu$ from~$m$). 
HQET starts with the infinite-mass limit, or static effective
theory~\cite{Eichten:1990zv,Grinstein:1990mj,Georgi:1990um}.
The eigenstates of this theory are independent of~$m$.
One can then develop the expansion in~$1/m$ of the underlying theory 
(QCD) around the infinite-mass limit, so that matrix elements are 
taken in the infinite-mass 
eigenstates~\cite{Eichten:1990vp,Luke:1990eg,Falk:1993wt}.

Our lattice method retains the logic and structure of the usual
application of HQET.
Lattice gauge theory with Wilson fermions has a stable heavy-quark 
limit~\cite{El-Khadra:1997mp}, in which the Isgur-Wise heavy-quark 
symmetries are prominent.
Indeed, the static limit is the same as for continuum~QCD.
Consequently, one may apply HQET directly to lattice gauge theory, to 
separate long- from short-distance physics~\cite{Kronfeld:2000ck}.
The key difference is that there are now \emph{two} short distances,
$1/m$ and the lattice spacing~$a$.
That does not run afoul of the assumptions of HQET; it means merely 
that the short-distance coefficients must be modified to depend on~$a$ 
as well as~$m$.
Then one may use HQET to develop heavy-quark expansions for lattice
observables.
The expansion for the rest mass~$M_1$ of a spin-$J$ meson
is~\cite{Kronfeld:2000ck}
\begin{equation}
	M_1 = m_1  +  \bar{\Lambda}_{\text{lat}}
		-     \frac{{\lambda_1}_{\text{lat}}}{2m_2} 
		- d_J \frac{{\lambda_2}_{\text{lat}}}{2m_{\cal B}}
		+ O(1/m^2),
	\label{eq:M1}
\end{equation}
where $m_1$, $1/2m_2$, and $1/2m_{\cal B}={z_{\cal B}}_{\text{lat}}/2m_2$ 
are the modified short-distance coefficients.
The rest mass and kinetic mass~$M_2$ are defined through the energy
\begin{equation}
	E(\bbox{p}) = M_1 + \frac{\bbox{p}^2}{2M_2} + \cdots
	\label{eq:E(p)}
\end{equation}
of a state with small momentum~$\bbox{p}$.
Because the lattice breaks Lorentz invariance, $M_2$ need not equal 
$M_1$, except asymptotically as $Ma\to 0$.
For quarks $m_1$ and $m_2$ are defined similarly in matching
calculations.

As $ma\to0$ lattice QCD becomes continuum QCD, so then $m_{1,2}\to m$
and~${z_{\cal B}}_{\text{lat}}\to z_{\cal B}$.
Owing to limitations in computer resources there are, however, no
lattice data available with $ma\ll 1$ and $m\gg\Lambda_{\text{QCD}}$.
The advantage of Eq.~(\ref{eq:M1}) is that it holds for general~$ma$,
as long as~$m_{2,\cal B}\gg\Lambda_{\text{QCD}}$.
One may, therefore, apply Eq.~(\ref{eq:M1}) to published data for~$M_1$.

Like their continuum-QCD counterparts, the
quantities~$\bar{\Lambda}_{\text{lat}}$, ${\lambda_1}_{\text{lat}}$,
and ${\lambda_2}_{\text{lat}}$ do not depend on the heavy-quark mass.
They are labeled with the subscript ``lat'' because the gluons and light 
quarks are also on the lattice.
Their lattice-spacing dependence can be separated from continuum QCD 
with Symanzik's formalism~\cite{Symanzik:1979ph}, which implies, for 
example,
\begin{equation}
	\bar{\Lambda}_{\text{lat}} = \bar{\Lambda} + a C_1 {\cal M}_1 + 
		a^2 C_2 {\cal M}_2 + \cdots,
	\label{eq:Symanzik}
\end{equation}
where the $C_i$ represent short-distance coefficients and the 
${\cal M}_i$ long-distance matrix elements in Symanzik's effective 
Lagrangian.
Equation~(\ref{eq:Symanzik}) is a good guide for extrapolating 
$a\to 0$ as soon as $\Lambda_{\text{QCD}}a\ll 1$.
Because lattice-spacing effects of the heavy quark are isolated in 
Eq.~(\ref{eq:M1}), it does not matter if $ma$ is not small.

Our method is to take Monte Carlo data for $M_1$ over a wide range of 
heavy quark masses, combine them with separate calculations of the 
short-distance coefficients, and perform fits to Eq.~(\ref{eq:M1}).
This is very simple for~${\lambda_2}_{\text{lat}}$:
\begin{equation}
	\case{1}{2}m_{\cal B}(M_{1B^*}-M_{1B}) = 
		{\lambda_2}_{\text{lat}},
	\label{eq:l2}
\end{equation}
with quark masses $m_2\gg\Lambda_{\text{QCD}}$
and with fixed $\mu$ (and~$a$).
For~$\bar{\Lambda}_{\text{lat}}$ and~${\lambda_1}_{\text{lat}}$ we 
consider the spin-averaged rest mass
$\bar{M}_1 := \case{1}{4}(3M_{1B^*}+M_{1B})$.
Then~${\lambda_2}_{\text{lat}}$ drops out, and Eq.~(\ref{eq:M1}) 
becomes
\begin{equation}
	\bar{M}_1	- m_1 = \bar{\Lambda}_{\text{lat}}
				-   \frac{{\lambda_1}_{\text{lat}}}{2m_2}.
	\label{eq:fit}
\end{equation}
Equation~(\ref{eq:fit}) is the crux of our analysis: we plot the 
combination on the left-hand side against~$(2m_2)^{-1}$, and a fit
to the mass dependence yields $\bar{\Lambda}_{\text{lat}}$ 
and~$-{\lambda_1}_{\text{lat}}$.
We repeat this procedure for several lattice spacings to take the 
continuum limit, guided by Eq.~(\ref{eq:Symanzik}).

To carry out the analysis one must calculate~$M_1$, for vector and
pseudoscalar mesons, and the short-distance coefficients $m_1$, $m_2$
and~$m_{\cal B}$.
For the coefficients we shall use perturbative QCD.
In lattice gauge theory
\begin{equation}
	m_X = m_X^{[0]} + \sum_{l=1}^\infty g_0^{2l}(1/a) m_X^{[l]},
	\label{eq:PT}
\end{equation}
where $g_0^2(1/a)$ is the bare coupling for a lattice with spacing~$a$.
For $m_1$ and $m_2$, Ref.~\cite{Mertens:1998wx} derived formulas to
relate the higher-order terms to the self energy and gave
the one-loop terms~$m_X^{[1]}$ for the lattice action used below.
For $m_{\cal B}$ only the tree-level term~$m_{\cal B}^{[0]}$
is known, so, for now, we cannot obtain a meaningful result
for~${\lambda_2}_{\text{lat}}$.

It is well-known that perturbation theory in~$g_0^2(1/a)$ converges
poorly.
Therefore, we re-express Eq.~(\ref{eq:PT}) in a renormalized
coupling, chosen with the Brodsky-Lepage-Mackenzie (BLM)
prescription~\cite{Brodsky:1983gc}.
For a coupling in scheme~$S$, we denote the BLM expansion
parameter~$g_S^2(q^*_S)$.
The BLM scale~$q^*_S$ is given by
\begin{equation}
	\log q^*_S = - \case{1}{2} b_S^{(1)} +
		\frac{\int d^4k\,\log k\,f(k)}{\int d^4k\,f(k)},
	\label{eq:qstar}
\end{equation}
where $k$ is the gluon momentum, and $f(k)$ is the integrand of the
quantity of interest, e.g., $\int d^4k\,f(k)=m_1$.
The constant $b_S^{(1)}$ is the $\beta_0$-dependent part of the
one-loop conversion from the arbitary scheme~$S$ to the ``$V$~scheme'',
namely
\begin{equation}
	\frac{(4\pi)^2}{g^2_S(q)} = \frac{(4\pi)^2}{g^2_V(q)} +
		\beta_0 b_S^{(1)} + b_S^{(0)} + O(g^2),
	\label{eq:bS}
\end{equation}
where for $n_f$ light quarks $\beta_0=11-2n_f/3$, 
and $b_S^{(0)}$ is independent of~$n_f$.
The $V$-scheme coupling $g^2_V(q)$ is defined so that the Fourier
transform of the heavy-quark potential reads $V(q)=-C_Fg^2_V(q)/q^2$.
Equation~(\ref{eq:qstar}) shows that the definitions of~$q^*$ in
Refs.~\cite{Brodsky:1983gc} and~\cite{Lepage:1993xa} are equivalent in
the $V$~scheme.%
\footnote{For convenience, we list some of the~$b_S^{(i)}$ here.
In the $V$~scheme $b_V^{(1)}=b_V^{(0)}=0$, by definition;
in the $\overline{\rm MS}$~scheme
$b_{\overline{\rm MS}}^{(1)}=5/3$,
$b_{\overline{\rm MS}}^{(0)}=-8$;
for the bare gauge coupling~\cite{Bode:2000sm}
$b_0^{(1)}=b_{\overline{\rm MS}}^{(1)}-6\pi K_1(1)=9.12637$,
$b_0^{(0)}=b_{\overline{\rm MS}}^{(0)}+2\pi[2d_{10}+33K_1(1)]=-16.1213$.
}

The purpose of the logarithmically weighted integral in
Eq.~(\ref{eq:qstar}) is to sum up into $g_S^2$ higher-order terms of
order~$g^2(\beta_0g^2)^{l-1}$, $l>1$, which with a foolish choice
of scale would be large.
The purpose of the constant is to make $g_S^2(q^*_S)$ independent of~$S$,
apart from contributions of order~$g^4(\beta_0g^2)^{l-2}$.
This is an advantage in matching calculations: it makes little
numerical difference whether one re-expands Eq.~(\ref{eq:PT})
in $g_0^2(q^*_0)$ or~$g_V^2(q^*_V)$.

In practice, we use $g_V^2(q^*_V)$, computed from the $1\times1$
Wilson loop and $g^2_V(3.40/a)$ as in Ref.~\cite{Lepage:1993xa}.
For~$m_1$ the BLM scale $q^*_V=q^*_1$ is now
available~\cite{Mertens:2000}.
Most of the loop correction to $m_2$ can be attributed
to~$m_1$, leaving an additional renormalization
factor~$Z_{m_2}$~\cite{Mertens:1998wx}.
The one-loop term is small~\cite{Mertens:1998wx}, but the BLM
scale~$q^*_2$ is not yet available.
So, for $Z_{m_2}$ we simply use $q^*_2=q^*_1\pm 20\%$, fully
correlated, and tolerate an extra uncertainty.

For lattice meson masses $M_1$ we select numerical data from recent
work on heavy-light pseudoscalar and vector
mesons~\cite{Aoki:1998ji,El-Khadra:1998hq,Becirevic:1999ua}.
The data are tabulated in Table~\ref{tab:data}.
\begin{table}[tp]
% \centering
\caption[tab:data]{Numerical and perturbative results used in this 
paper.
The first column cites the source of the numerical data.
The second column includes the plotting symbol used in all figures.
Statistical errors are given for $a^{-1}$ and $\bar{M}_1$;
systematic (perturbative) errors for~$m_2$.
Perturbative results are from input data to the numerical calculations 
and Refs.~\cite{Mertens:1998wx,Mertens:2000}.}
\begin{tabular}{ccllll}
	 Ref.\  & $a^{-1}$~(GeV)
	        & \multicolumn{1}{c}{$\bar{M}_1a$} 
	        & \multicolumn{1}{c}{$m_1a$}
	        & \multicolumn{1}{c}{$m_2a$}
	        & \multicolumn{1}{c}{$q^*_1a$} \\
	\hline
	%
%% kek6.3
\protect\cite{Aoki:1998ji}
        &3.35(2)        &1.731(3)	 &1.560	 &2.761(54)	 &0.91 \\
        &octagons       &1.301(3)	 &1.122	 &1.567(21)	 &0.90 \\
        &               &0.946(2)	 &0.761	 &0.903(7)	 &0.85 \\
        &               &0.789(2)	 &0.602	 &0.674(4)	 &0.80 \\
        &               &0.667(2)	 &0.477	 &0.514(2)	 &0.75 \\
        &               &0.589(2)	 &0.398	 &0.420(1)	 &0.70 \\
        &               &0.523(2)	 &0.331	 &0.343(1)	 &0.65 \\
%     	       
%% kek6.1	       
        &2.50(2)        &2.147(4)	 &1.943	 &4.216(132)	 &0.90 \\
        &squares        &1.611(3)	 &1.399	 &2.217(49)	 &0.91 \\
        &               &1.183(3)	 &0.961	 &1.229(17)	 &0.88 \\
        &               &0.978(3)	 &0.750	 &0.880(9)	 &0.84 \\
        &               &0.845(3)	 &0.613	 &0.686(5)	 &0.80 \\
        &               &0.749(2)	 &0.514	 &0.558(3)	 &0.76 \\
        &               &0.676(2)	 &0.438	 &0.466(2)	 &0.71 \\ \hline
%
%% fnal6.1	       
\protect\cite{El-Khadra:1998hq}
        &2.50(2)        &2.557(7)	 &2.364	 &6.722(234)	 &0.88 \\
        &crosses        &1.403(14)	 &1.200	 &1.716(29)	 &0.90 \\
        &               &0.726(10)	 &0.504	 &0.545(3)	 &0.75 \\
%     	       
%% fnal5.9	       
        &1.77(1)        &2.665(6)	 &2.422	 &6.773(373)	 &0.88 \\
        &diamonds       &1.663(4)	 &1.402	 &2.163(66)	 &0.91 \\
        &               &0.964(4)	 &0.677	 &0.770(9)	 &0.81 \\
        &               &0.876(4)	 &0.582	 &0.642(6)	 &0.77 \\
%     	           	       
%% fnal5.7	       
        &1.16(1)        &2.829(6)	 &2.535	 &6.735(806)	 &0.88 \\
        &fancy          &2.345(6)	 &2.037	 &4.067(381)	 &0.90 \\
        &squares        &1.935(6)	 &1.612	 &2.599(179)	 &0.91 \\
        &               &1.489(5)	 &1.139	 &1.498(63)	 &0.89 \\
        &               &1.274(5)	 &0.917	 &1.112(34)	 &0.85 \\ \hline
%     	       
%% ape6.2	       
\protect\cite{Becirevic:1999ua}
        &2.90(2)        &0.958(7)	 &0.748	 &0.883(8)	 &0.84 \\
        &fancy          &0.849(6)	 &0.636	 &0.719(5)	 &0.81 \\
        &diamonds       &0.762(6)	 &0.548	 &0.602(3)	 &0.78 \\
        &               &0.670(5)	 &0.454	 &0.485(2)	 &0.73 \\

\end{tabular}
\label{tab:data}
\end{table}
For uniformity, the value in physical units of the lattice spacing~$a$
is defined according to the suggestion of Sommer~\cite{Sommer:1994ce}.
(It gives the same numerical result as the 1P-1S splitting of
charmonium.)
The lattice spacing varies by a factor of nearly~3, allowing us the
take the continuum limit as guided by Eq.~(\ref{eq:Symanzik}).
All data sets are in the quenched approximation, which omits the
back-reaction of light quarks on the gluons and partly compensates
the omission by implicit shifts in the bare couplings.
Light quarks have the Sheikholeslami-Wohlert
action~\cite{Sheikholeslami:1985ij},
to minimize discretization effects on the light quark.
In most data sets, the (physical) quark mass spans a range from near
charm to slightly above beauty, allowing us to examine the mass
dependence of Eq.~(\ref{eq:fit}).

\begin{figure}[tbp]
\centering
\epsfxsize=0.48\textwidth \epsfbox{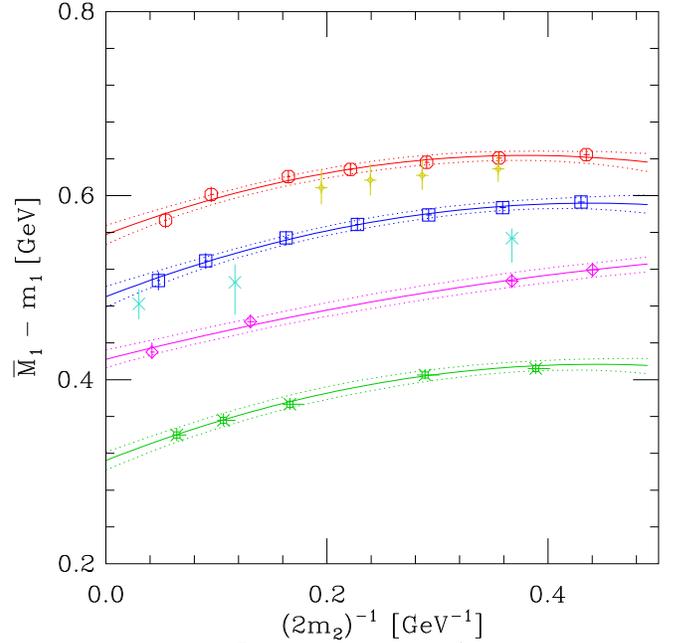}
\caption[fig:data]{Plot of $\bar{M}_1-m_1$ vs.\ $(2m_2)^{-1}$.
The key for the plotting symbols is given in Table~\ref{tab:data}.
For clarity the error envelopes for the crosses and fancy diamonds
are not shown.}
\label{fig:data}
\end{figure}

Figure~\ref{fig:data} plots $\bar{M}_1-m_1$ vs.\ $(2m_2)^{-1}$.
The vertical error bars reflect statistical uncertainties only,
and the horizontal error bars reflect these and the variation
in~$q^*_2$.
There is noticeable curvature, which is not surprising because the
data reach masses below the charmed quark mass.
We handle the curvature in two ways.
First, we fit linearly the subset of data with $m_2\ge2.5$~GeV.
Second, we extend Eq.~(\ref{eq:fit}) to
order~$1/m^2$~\cite{Kronfeld:2000ck}:
\begin{equation}
	\bar{M}_1	- m_1 = \bar{\Lambda}_{\text{lat}}
				-   \frac{{\lambda_1}_{\text{lat}}}{2m_2}
				+      \frac{{\rho_1}_{\text{lat}}}{4m_D^2}
				-  \frac{{{\cal T}_1}_{\text{lat}}
				+        {{\cal T}_3}_{\text{lat}}}{(2m_2)^2},
	\label{eq:barM1rhoT}
\end{equation}
where $1/4m_D^2$ is the short-distance coefficient of the Darwin term,
and ${\rho_1}_{\text{lat}}$, ${{\cal T}_1}_{\text{lat}}$, and
${{\cal T}_3}_{\text{lat}}$ are matrix elements of higher-dimension
terms~\cite{Mannel:1994kv,Bigi:1995ga,Gremm:1997df}, with the notation
of Ref.~\cite{Gremm:1997df}, for gluons and light quarks on a lattice.
The~$1/m^2$ terms are important for smaller masses, where
$m_D\approx m_2$ within the precision available.
Thus, only one unknown is needed to model the curvature.

We take the second method as our standard and use the first
for comparison.
The solid curves in Fig.~\ref{fig:data} are the best fit to
Eq.~(\ref{eq:barM1rhoT}).
We use the bootstrap method to propagate the
underlying uncertainties through the fit.
In this way we account fully (partially) for correlations in the data
from Ref.~\cite{El-Khadra:1998hq}
(Refs.~\cite{Aoki:1998ji,Becirevic:1999ua}).
The dotted lines show the error envelopes of the fits;
they hug the best fit in the region of interpolation and
flare out in the region of extrapolation.

As expected, $\bar{\Lambda}_{\text{lat}}$ and~${\lambda_1}_{\text{lat}}$
depend on the lattice spacing~$a$.
For the data sets used, the coefficient $C_1$ in Eq.~(\ref{eq:Symanzik})
is of order~$\alpha_s$ and the coefficient $C_2$ is of order~1.
Asymptotically, the former dominates, so we fit
$\bar{\Lambda}_{\text{lat}}$ linearly in $a$
to take the continuum limit.
The slope, $C_1{\cal M}_1$, is somewhat large for a quantity of order
$\alpha_s\Lambda_{\text{QCD}}^2$, so we also consider fits linear
in~$a^2$.
The~$\chi^2/$dof is smaller for the fit linear in~$a$, so we take it
for our central value and take the other to quote a systematic error.
In future work, one should tune the light quark action so that $C_1$
is of order~$a$~\cite{Luscher:1997ug}; then the extrapolation
Ansatz would be unambiguous.

\begin{figure}[tbp]
\centering
\epsfxsize=0.48\textwidth \epsfbox{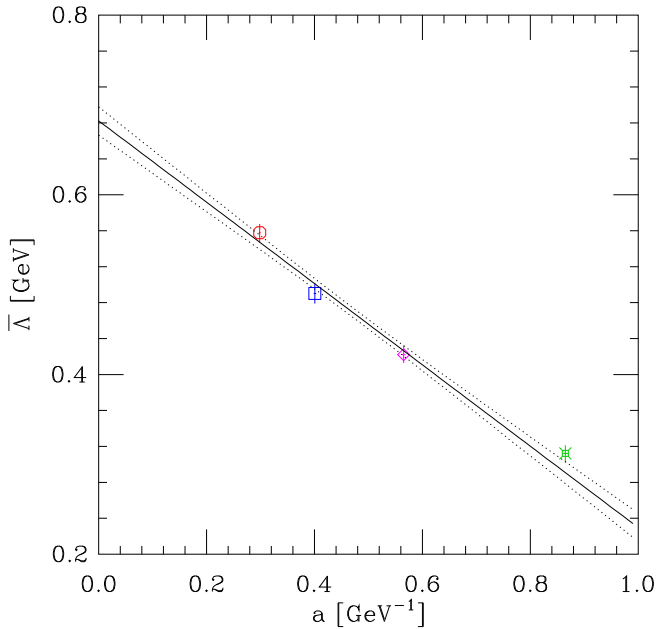}
\caption[fig:clL]{Continuum limit of~$\bar{\Lambda}$.}
\label{fig:clL}
\end{figure}

Figure~\ref{fig:clL} plots $\bar{\Lambda}_{\text{lat}}$ vs.~$a$.
The error bars are from the bootstrap of the fit described above.
From now on we discard the data sets denoted in Fig.~\ref{fig:data}
by crosses and fancy diamonds.
Their error bars are very large: the crosses have too few points and
the mass range of the fancy diamonds is too small.
$\bar{\Lambda}_{\text{lat}}$~exhibits significant dependence on~$a$;
in this case, it would have been misleading to determine
$\bar{\Lambda}$ with data at only one lattice spacing.

\begin{figure}[tbp]
\centering
\epsfxsize=0.48\textwidth \epsfbox{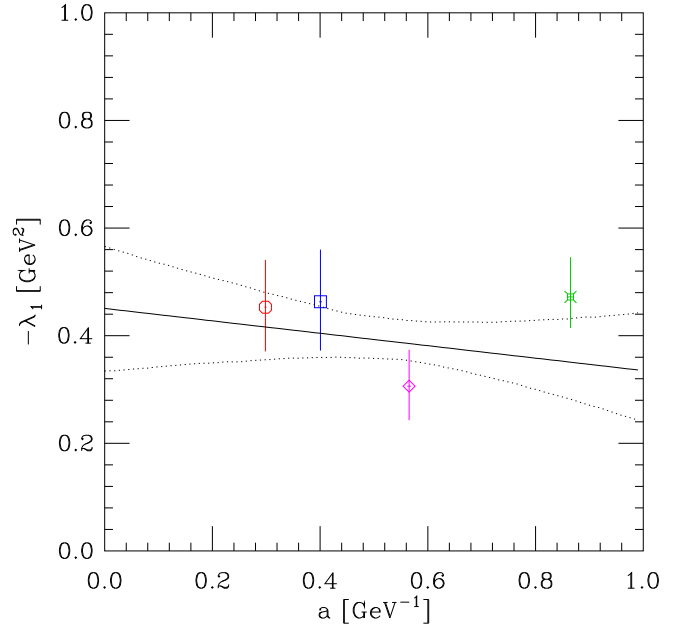}
\caption[fig:cl1]{Continuum limit of~$\lambda_1$.}
\label{fig:cl1}
\end{figure}

Figure~\ref{fig:cl1} plots ${\lambda_1}_{\text{lat}}$ vs.~$a$.
The error bars are again from the bootstrap of the mass fit.
In this case, lattice spacing effects are smaller than other
uncertainties, and it does not matter whether we take the continuum
limit with a fit to $a$ or to~$a^2$.

\begin{figure}[tbp]
\centering
\epsfxsize=0.48\textwidth \epsfbox{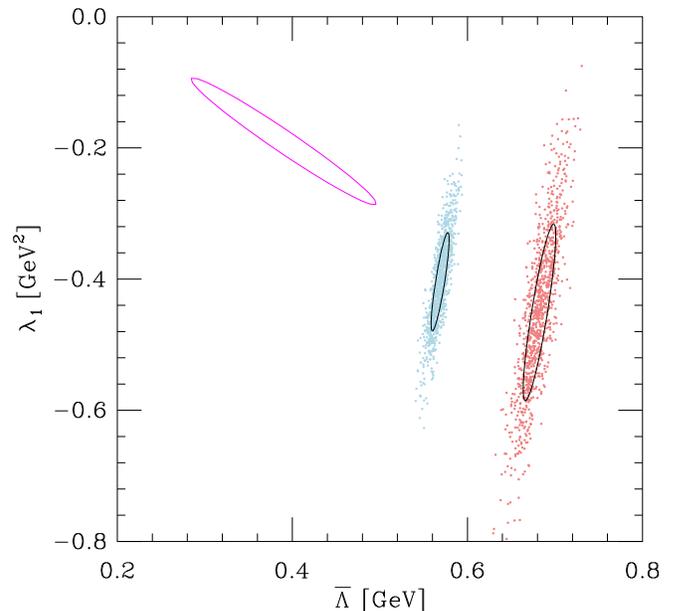}
\caption[fig:cor]{Correlation of our results for~$\bar{\Lambda}$ 
and~$\lambda_1$ from two analyses of the continuum limit.
Dark grey (red) points are the standard analysis, quadratic in $1/2m_2$
and linear in~$a$.
Light grey (blue) points are quadratic in $1/2m_2$ and linear
in~$a^2$, yielding smaller~$\bar{\Lambda}$.
The ellipse in the upper left is the result of
Ref.~\cite{Gremm:1996yn}.}
\label{fig:cor}
\end{figure}

The results exhibit a strong correlation in the
$\bar{\Lambda}$-$\lambda_1$ plane, as shown in Fig.~\ref{fig:cor}.
The points show the scatter from the bootstrap method.
The ellipses surround 68\% of the points.
Dark grey (red) points show the standard analysis, with fits
quadratic in $1/2m_2$ and linear in~$a$.
Light grey (blue) points show the analysis with continuum
extrapolation linear in~$a^2$, yielding smaller~$\bar{\Lambda}$.
The results from four different Ans\"atze for fitting are tabulated in
Table~\ref{tab:fit}.
\begin{table}[tp]
% \centering
\caption[tab:fit]{Numerical results for four different fit Ans\"atze.
The column labeled $\rho$ gives the normalized coefficient of
correlation.}
\begin{tabular}{lccc}
	\multicolumn{1}{c}{fit}	&
		$\bar{\Lambda}$ (GeV) & $-\lambda_1$ (GeV$^2$) & $\rho$ \\
	\hline
	Eq.~(\ref{eq:barM1rhoT}),  $a$  &
		$0.68  \pm 0.02$ & $0.45  \pm 0.12$ & 0.869 \\
	Eq.~(\ref{eq:fit}),        $a$  &
		$0.67^{+0.01}_{-0.02}$ & $0.44  \pm 0.11$ & 0.852 \\
	Eq.~(\ref{eq:barM1rhoT}), $a^2$ &
		$0.57  \pm 0.01$ & $0.40  \pm 0.07$ & 0.860 \\
	Eq.~(\ref{eq:fit}),       $a^2$ &
		$0.57  \pm 0.01$ & $0.41  \pm 0.08$ & 0.871 \\
\end{tabular}
\label{tab:fit}
\end{table}

Clearly the choice of lattice-spacing extrapolation dominates the
uncertainties of Monte Carlo statistics and~$q^*_2$,
which are propagated carefully through the fits.
To account for this, we take central value for $\bar{\Lambda}$ from
the standard analysis, but we extend the error bar to encompass the
full range suggested by the $a^2$ fit.
On the other hand, the standard fit gives an error bar for~$-\lambda_1$
that covers the range of the other fits, so we simply use it.
With these considerations we find,
\begin{eqnarray}
	\bar{\Lambda} & = & 
		0.68^{+0.02}_{-0.12}~\text{GeV},
	\label{eq:Lambda_lin}  \\
	\lambda_1     & = &
 		-(0.45 \pm 0.12)~\text{GeV}^2.
	\label{eq:lambda1_lin}
\end{eqnarray}
The standard fit also yields an estimate of dimension-three combination
${\cal T}_1+{\cal T}_3-\rho_1=0.51\pm 0.22~\text{GeV}^3$.

The orientation of the ellipses from our method is
roughly orthogonal to that found from moments of the lepton energy
spectrum~\cite{Gremm:1996yn,Gremm:1997gg} or the hadronic invariant
mass spectrum~\cite{Falk:1998jq} of inclusive~$B$ decay.
For illustration, the former is shown in Fig.~\ref{fig:cor} as well.

There are two uncertainties that we cannot yet address fully.
One is the effect of finite volume on~$M_1$.
Studies of the volume dependence of heavy-light
systems~\cite{Alexandrou:1994ti,Duncan:1995uq} suggest that
finite-volume effects are negligible compared to our other
uncertainties.
A~more serious uncertainty arises because the numerical data were
generated in the quenched approximation.
One may expect that the shift in~$\bar{\Lambda}$ owing to quenching is
small, for the same reason that the shift in the heavy-quark mass is
small~\cite{Davies:1994pz}.
A~qualitative way of estimating the effect of quenching is to check
other, similar observables.
With Sommer's definition of~$a$ one finds discrepancies in~$m_\rho$
of around $-10$ percent, suggesting that $\bar{\Lambda}$ could be 10
percent smaller, and~$\lambda_1$ 20 percent larger, than quoted here.

We do not quote an uncertainty from the perturbative calculation of
the short-distance effects.
Because HQET, as customarily applied, is defined with a perturbative
renormalization scheme, any application suffers from such
uncertainties.
Our results for $\bar{\Lambda}$ and $\lambda_1$ can be used
consistently with the pole mass in next-to-leading order, BLM-improved
phenomenology.
In such an application a single uncertainty from truncating
perturbative QCD should be quoted.
Indeed, because the pole mass has large higher-order contributions,
so does~$\bar{\Lambda}$, but in a physical application the large
terms cancel.
If next-to-next-to-leading accuracy is required, then the analysis
presented here must be repeated with (as yet uncalculated) two-loop
short-distance coefficients.

Our central value for $\bar{\Lambda}$ is somewhat larger than 
those from QCD sum rules~\cite{Bagan:1992sg}, but taking the
uncertainties into account, there is no inconsistency.
Our result for $\lambda_1$ agrees with some sum-rule estimates,
but not others~\cite{Ball:1994xv}.
It is not clear what to make of the discrepancies in sum rules.
Our uncertainties are reducible, and below we identify ways to improve
the numerical data that go into our analysis.

In the past, there have been attempts to calculate the lambdas in a
discretization of the infinite-mass limit~\cite{Crisafulli:1995pg}.
This method faces two difficulties.
First, it yields the lambdas in a lattice renormalization scheme, and
the results must be converted to the continuum schemes in common use.
The conversion must deal with power-law
divergences~\cite{Maiani:1992az}.
Second, it identifies the HQET separation scale~$\mu$ with the
ultraviolet cutoff $\pi/a$ of the gluons, so it is hard to take the
continuum limit.
Our method circumvents these obstacles by formulating HQET as an
effective field theory to describe sets of (lattice) data.
In this way HQET obtains its own scale~$\mu$ and the second problem
does not arise.
The first problem arises from taking $m\to\infty$ with $a$ fixed.
Our method sidesteps it by fitting the mass dependence in the
regime $m_1a\lesssim2$, and, since $m\gg\Lambda_{\text{QCD}}$, HQET
identifies the fit parameters with~$\bar{\Lambda}_{\text{lat}}$,
${\lambda_1}_{\text{lat}}$, and ${\lambda_2}_{\text{lat}}$.

In this paper, we have presented a new way to determine
$\bar{\Lambda}$, $\lambda_1$, and $\lambda_2$.
Using numerical data in the literature, we have shown that it is
feasible to carry out the necessary fit in quark mass and extrapolation
in lattice spacing to obtain encouraging results.
Systematic uncertainties in the mass extrapolation might be improved
using the hopping-parameter expansion~\cite{Henty:1992cw}, to create
a continuous range of heavy-quark mass.
With small enough statistical errors and a wide enough range of data,
it might be possible also to extract the dimension-three quantities
$\rho_i$ and~${\cal T}_i$, although that task requires the calculation
of several additional short-distance coefficients.
Similarly, systematic uncertainties in the lattice-spacing extrapolation
could be improved by adjusting the light quarks' action so that $C_1$ in
Eq.~(\ref{eq:Symanzik}) is rendered of order $a$~\cite{Luscher:1997ug}.
Finally, our methods could be applied to full QCD, once such data sets
have been generated, to obtain truly \emph{ab initio} results.

% \acknowledgments
We thank Shoji Hashimoto for sending us data used, but not tabulated,
in Ref.~\cite{Aoki:1998ji}.
Fermilab is operated by Universities Research Association Inc.,
under contract with the U.S. Department of Energy.

\end{document}